\documentclass[aps,pre,twocolumn,showpacs,superscriptaddress]{revtex4}
\usepackage{amsmath,amssymb}
\usepackage{graphicx}% Include figure files
\newcommand{\sech}{\mbox{sech}}
\newcommand{\cn}{\mbox{cn}}

\begin{document}

%\preprint{}

\title{Loss of stability of a solitary wave through exciting a cnoidal wave on a Fermi-Pasta-Ulam ring}


%\author{Zongqiang Yuan}
%\thanks{}
%\affiliation{BNU}

\author{Zongqiang Yuan}
\affiliation{Department of Physics and the Beijing-Hong Kong-Singapore Joint Centre for Nonlinear and Complex Systems (Beijing), Beijing Normal University, Beijing 100875, China}
\author{Jun Wang}
\affiliation{Key Laboratory of Enhanced Heat Transfer and Energy Conservation, Ministry of Education, College of Environmental and Energy Engineering, Beijing University of Technology, Beijing 100124, China}
\author{Min Chu}
\affiliation{Department of Physics and the Beijing-Hong Kong-Singapore Joint Centre for Nonlinear and Complex Systems (Beijing), Beijing Normal University, Beijing 100875, China}
\author{Guodong Xia}
\affiliation{Key Laboratory of Enhanced Heat Transfer and Energy Conservation, Ministry of Education, College of Environmental and Energy Engineering, Beijing University of Technology, Beijing 100124, China}
\author{Zhigang Zheng}
\email[]{zgzheng@bnu.edu.cn}
\affiliation{Department of Physics and the Beijing-Hong Kong-Singapore Joint Centre for Nonlinear and Complex Systems (Beijing), Beijing Normal University, Beijing 100875, China}


%\date{\today}

\begin{abstract}
The spatiotemporal propagation behavior of a solitary wave is investigated on a Fermi-Pasta-Ulam ring. We observe the emergence of a cnoidal wave excited by the solitary wave. The cnoidal wave may coexist with the solitary wave for a long time associated with the periodic exchange of energy between these two nonlinear waves. The module of the cnoidal wave, which is considered as an indicator of the nonlinearity, is found to oscillate with the same period of the energy exchange. After the stage of coexistence, the interaction between these two nonlinear waves leads to the destruction of the cnoidal wave by the radiation of phonons. Finally, the interaction of the solitary wave with phonons leads to the loss of stability of the solitary wave.
\end{abstract}



%\keywords{}

\maketitle

\section{introduction}
Nonlinear waves can trace their history back to Russell's discovery of "the wave of translation" (now known as solitary wave or soliton) on the Union Canal in Scotland in 1834~\cite{Russell1844,Whitham1974}.
About 60 years ago Fermi, Pasta, and Ulam (FPU) introduced the FPU model to investigate the energy equipartition problem and the ergodic hypothesis in statistical physics~\cite{Fermi1955}.
The attempt to resolve the mystery of the FPU recurrence has led to the rediscovery of solitons~\cite{Zabusky1965}. It is now widely accepted that solitary waves are of great importance in diverse areas of science and technology~\cite{Kivshar1989,Sen2008,Kartashov2011,Dauxois2006,Akhmediev2005,Akhmediev2008}.
In addition to the solitary waves, some other kinds of nonlinear waves have also been studied in a variety of systems, such as intrinsic localized modes (or discrete breathers)~\cite{Sievers1988,Takeno1988,MacKay1994,Flach1998,Flach2008,Sato2006} and cnoidal waves~\cite{Driscoll1976,Shultz1997,Kartashov2003,Jeng2009,Friesecke2012}.

The remarkable stability of the solitary waves is one of the reasons why they have attracted much attention. An extensive literature on the subject of stability of the solitary waves developed.
Usually, stability analysis of a solitary wave is considered with respect to the infinitesimally perturbation of the type $\exp(iQx+\Omega t)$ and the dispersion relation $\Omega=\Omega(Q)$ is derived. The solitary wave is said to be stable with respect to the perturbation if the real part of $\Omega$ is negative, while, the solitary wave is unstable if the real part of $\Omega$ is positive. Actually, for the unstable case, it is possible that the infinitesimal perturbation grows into a finite perturbation which is then stabilized by the nonlinearity~\cite{Berryman1976}.
The stability of solitary wave solutions of the Korteweg-de Vries (KdV) equation was worked out by Benjamin~\cite{Benjamin1972}, while the asymptotic stability of the solitary waves of the KdV and a class of generalized KdV equations was done by Pego and Weinstein~\cite{Pego1994}.
Under the long-wavelength approximations, the FPU-$\alpha$ and FPU-$\beta$ lattices result, respectively, in the KdV and modified KdV equations.
The existence theorem for the solitary waves on the FPU lattices is established in Ref.~\cite{Friesecke1994}.
The stability of solitary wave solutions on the FPU lattices at low energy was proven by Friesecke and Pego~\cite{Friesecke1999,Friesecke2002,Friesecke2004,Friesecke2004a}.
However, there has been very little work done on the stability of solitary waves at high-energy level.

In the present paper, we construct a solitary wave moving on a FPU-$\beta$ ring at high-energy level and study its spatiotemporal propagation behavior.
We observed the loss of stability of the solitary wave.
Three stages can be identified in the process of the loss of stability of the solitary wave.
In the first stage, the solitary wave excites a cnoidal wave due to the modulational instability of the system. These two kinds of nonlinear waves coexist associated with the periodic exchange of energy between them. The module of the cnoidal wave (i.e., the module of the corresponding Jacobian elliptic function) oscillates with the same period of the energy exchange.
In the second stage, the interaction of the cnoidal wave with the solitary wave leads to the radiation of phonons and the destruction of the cnoidal wave. In the third stage, the solitary wave loses stability due to the interaction with phonons. The process of phonon radiation can be shown directly by the spectral energy density (SED) method recently developed by Thomas \textit{et al.}~\cite{Thomas2010}.

\section{model and method}
The Hamiltonian of the FPU-$\beta$ model can be written as
\begin{equation}
\begin{aligned}
H &= \sum _{n}\Big [\frac{p_n^2}{2} + V(u_{n+1},u_n)\Big],\\
V(u_{n+1},u_n)&=\frac{k}{2}\Bigl(u_{n+1}-u_n\Bigr)^2 + \frac{\beta}{4}\Bigl(u_{n+1}-u_n \Bigr)^4,
\label{DefineHamiltonian}
\end{aligned}
\end{equation}
where $p_n$ and $u_n$ denote the momentum and the displacement from the equilibrium position of the $n$th particle, respectively. In the absence of the quartic term, i.e., $\beta = 0$, the above Hamiltonian reduces to a one-dimensional harmonic lattice, which is integrable and can be analytically solved. The presence of the anharmonic terms breaks the integrability and brings forth kinds of nonlinear effects.

The momentum excitation method is widely employed in the studies of modes excited on the lattices~\cite{Zavt1993,Hu2000,Rosas2004,Zhao2005}.
To construct a solitary wave moving on a FPU-$\beta$ ring, we firstly impart a momentum excitation $\tilde p$ to the left end particle of an initially quiescent chain consisting of $N_0+N$ particles (-$N_0$+1, -$N_0$+2, ..., -1, 0, 1, ..., N) with free boundary condition. If the momentum excitation is large enough, a solitary wave could be excited at the left-hand side accompanied by a low-amplitude wave called the tail~\cite{Hu2000}. The solitary wave may isolate itself naturally from the tail as it moves coherently along the chain faster than the tail. When the solitary wave arrives at particle $N/2$, we reconnect particle $1$ to particle $N$ in order to form a $N$-particle ring. As long as $N_0$ is large enough and $N_0 \gg N$, the tail will be wiped away and a solitary wave moving on a FPU-$\beta$ ring is constructed. The fourth-order symplectic method is employed in order to solve the dynamics of the FPU-$\beta$ model as a high-dimensional Hamilton dynamical system~\cite{QinMeng-zhao1991}. In the present paper, $\tilde p=10$, $N_0=5000$, $N=100$, $k=0.5$, $\beta=0.126$ and integration time step $dt=0.001$ throughout the simulation (unless stated otherwise).

\section{propagation behavior of the solitary wave}
To understand the dynamics of the solitary wave, we resort to the evolution of the energy of the low-amplitude wave initially excited on the chain or later on the ring. For the lattice system (chain or ring) we are studying here, the energy of the low-amplitude wave is defined as the residual energy of the solitary wave. Due to the spatial localization of the solitary wave, one can write the low-amplitude wave's energy as
\begin{equation}
\label{DefineEw}
 E_W = E(t) - \sum_{n=n_c-n_b}^{n_c+n_b} E_n(t),
\end{equation}
where $E(t)$ is the instantaneous total energy of the lattice system. The local energy $E_n(t)$ of the $n$th particle is defined as $E_n = \frac{p_n^2}{2} + \frac{1}{2} V(u_{n+1},u_n) + \frac{1}{2} V(u_n,u_{n-1})$. $n_c(t)$ is the center position of the solitary wave at time $t$.  $n_b$ denotes the number of the left/right neighboring particles of the center particle of the solitary wave packet. Numerically $n_b=5$ is enough due to the energy localization of the solitary wave.
The increase of $E_W$ on the ring corresponds to the decrease of the solitary wave energy and vice versa considering that the total energy of the system is conserved.

\begin{figure}
\includegraphics[width=\linewidth]{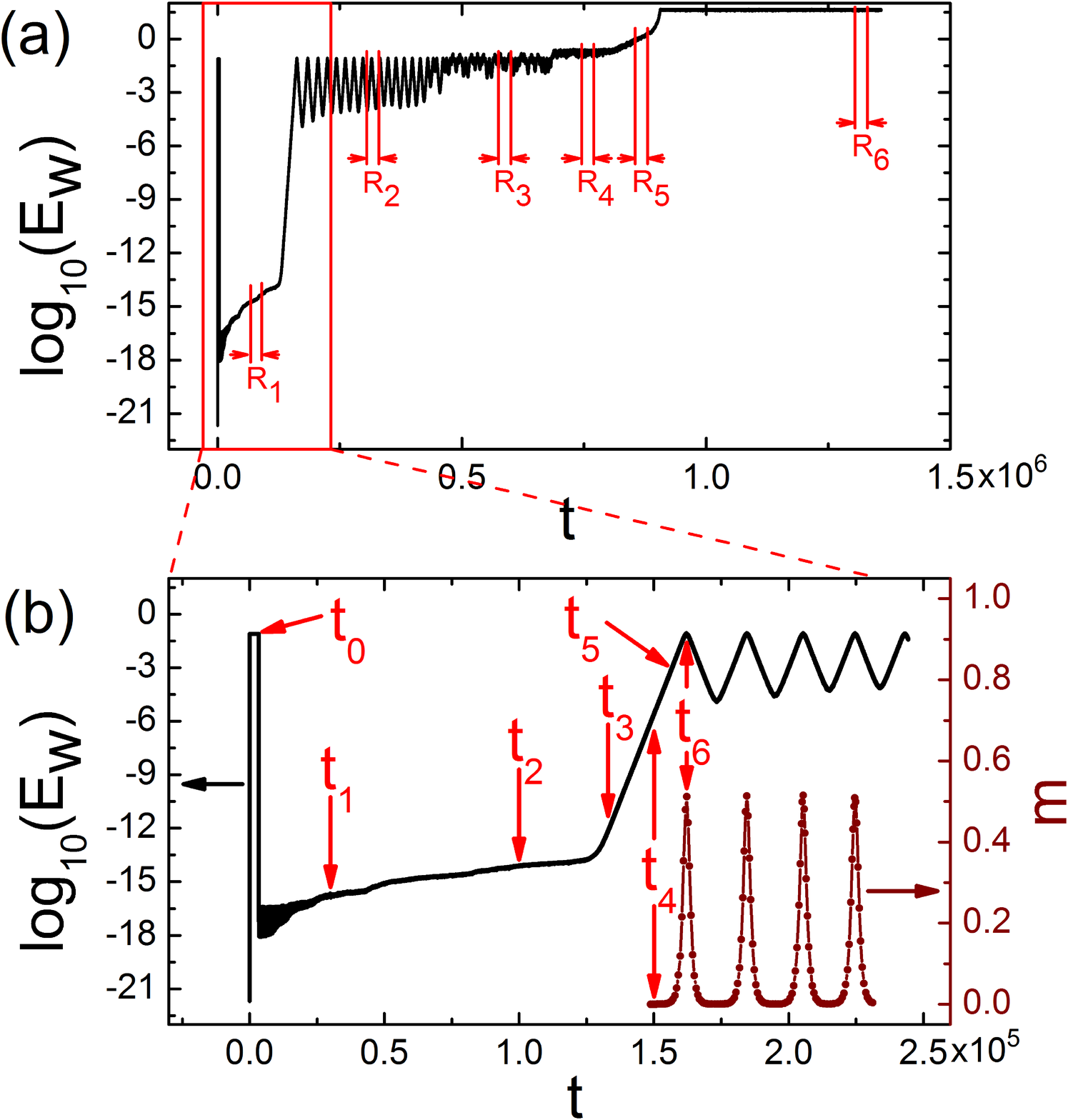}
\caption{\label{LAWEnergy} (Color online) (a) Time evolution of the energy of the low-amplitude wave. For clarity, part of (a) is enlarged in the left-hand side of (b). Right-hand side of (b) presents the time evolution of the module of the cnoidal wave.}
\end{figure}

\begin{figure}
\includegraphics[width=\linewidth]{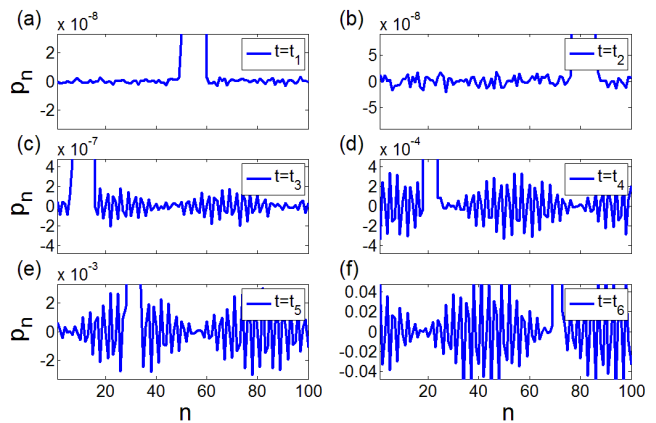}
\caption{\label{Momentum} (Color online) Momentum distribution profiles among the particles for different moments $t_1$-$t_6$ indicated in Fig.~\ref{LAWEnergy}(b). Vertical axes have been adjusted appropriately to get clear observations of the low-amplitude wave.}
\end{figure}

Figure~\ref{LAWEnergy}(a) presents the time evolution of $E_W$. It is clearly shown that the solitary wave moving on the ring may lose stability in despite of the lack of initially excited low-amplitude wave and boundary effects. To explore the underlying mechanism, we are now concerned with the evolution behavior of $E_W$ on a relatively short time scale as shown in the left-hand side of Fig.~\ref{LAWEnergy}(b). The fast increase of $E_W$ at the very beginning indicates the separation between the solitary wave and the tail. The moment $t_0$ corresponds to the construction of a solitary wave moving on a ring by connecting particles 1 and N.

It is instructive to analyze the momentum evolution behavior of the particles on the ring as depicted in Fig.~\ref{Momentum}. As will be shown below, Figure~\ref{Momentum} reveals the formation of a cnoidal wave and the coexistence of the cnoidal wave with the solitary wave.
The equations of motion corresponding to the Hamiltonian (\ref{DefineHamiltonian}) are
\begin{eqnarray}
\label{EquationOfMotionCopy}
\ddot u_n&=&k(u_{n+1}-u_n)+k(u_{n-1}-u_n) \nonumber\\
&&+\beta(u_{n+1}-u_n)^3+\beta(u_{n-1}-u_n)^3,
\end{eqnarray}
where the dots represent the time derivative $\partial/\partial t$.
For analytical consideration, we use multiple scale analysis presenting $u_n$ as the multiplication of harmonic oscillation and smooth envelope function~\cite{Khomeriki2002}
\begin{equation}
\label{Varphi2ToParticlePosition}
 u_n = \frac{\epsilon}{2} \varphi(\xi,\tau) e^{i(qn-\omega t)} + c.c.,
\end{equation}
where c.c. stands for complex conjugation. The appropriate slow variables are defined as $\xi = \epsilon(n-v t), \ \ \tau=\epsilon^2 t$, where $\epsilon$ is a formal small parameter indicating the smallness or slowness of the variables before which it appears. $v$ is the velocity of the cnoidal waves as will be shown below.
Substituting (\ref{Varphi2ToParticlePosition}) into Eq.~(\ref{EquationOfMotionCopy}) and neglecting the higher harmonics in a rotating wave approximation, in the first order over $\epsilon$ a well known dispersion relation for linear excitations in the FPU model is obtained
\begin{equation}
\label{NLSDeductionCompareEpsilon1}
\omega=\sqrt{2k(1-\cos q)}.
\end{equation}
For the second order over $\epsilon$, we have
\begin{equation}
\label{NLSDeductionCompareEpsilon2}
v=\frac{k \sin q}{\omega}.
\end{equation}
We get the nonlinear Schr\"{o}dinger (NLS) equation for the envelope function $\varphi$ in the third order over $\epsilon$
\begin{equation}
\label{NLSEquationFromFPUbeta}
i \frac{\partial \varphi}{\partial \tau} - \frac{\omega}{8}\frac{\partial^2 \varphi}{\partial \xi^2} - \frac{3\beta \omega^3}{8k^2}|\varphi|^2 \varphi=0.
\end{equation}
Following the standard process in Ref.~\cite{Whitham1974}, a subfamily of exact periodic solutions in the form of cnoidal waves [expressed in terms of the Jacobian elliptic functions $\cn(x,m)$] for the above NLS equation can be derived as
\begin{eqnarray}
\label{CnoidalWaveSolutionOfNLSEquationFromFPUbeta}
 \varphi &=& \sqrt{h_3} \cn \Big(\sqrt{h_3-h_1}\sqrt{\frac{3\beta \omega^2}{2k^2}}\xi-C_3,m\Big) \nonumber\\
 &&\times \exp\Big\{-i\Big[\big(h_3+h_1\big)\frac{3\beta \omega^3}{16k^2}\tau -C_4 \Big]\Big\},
\end{eqnarray}
where $h_1$ and $h_3$ are parameters which satisfy the relation $h_1<0<h_3$, $m=\frac{h_3}{h_3-h_1}$ is the module of the Jacobian elliptic function, and $C_3$ and $C_4$ are the constants of integration.
Physically, the elliptic parameter $m$ may be viewed as a fair indicator of the nonlinearity with the linear limit being $m \to 0$ and the extreme nonlinear limit being $m \to 1$.
If $|h_1| \ll h_3$, then $\cn(x,m \to 1) \to \sech(x)$ and the well-known envelope soliton solutions to the NLS equation~(\ref{NLSEquationFromFPUbeta}) are recovered. If $|h_1| \gg h_3$, then $\cn(x,m \to 0)\to \cos(x)$ with vanishingly small amplitude.
Substituting (\ref{CnoidalWaveSolutionOfNLSEquationFromFPUbeta}) into (\ref{Varphi2ToParticlePosition}), with the help of $\xi = \epsilon(n-v t)$, $\tau=\epsilon^2 t$ and $\epsilon=1$, we have
\begin{eqnarray}
\label{ParticlePositionTheory}
 u_n &=&\sqrt{h_3} \cn \Big[\sqrt{h_3-h_1}\sqrt{\frac{3\beta \omega^2}{2k^2}}(n -v t)-C_3,m\Big] \nonumber\\
 &&\times \cos\Big[qn-\Big(\omega+\frac{3\beta \omega^3}{16k^2}\big(h_3+h_1\big)\Big)t -C_4 \Big].
\end{eqnarray}
After the time derivative of $u_n$ and assuming that $h_1$ and $h_3$ are very small, the analytic formula for the cnoidal wave solutions of the FPU-$\beta$ model (\ref{DefineHamiltonian}) is obtained
\begin{eqnarray}
\label{ParticleMomentumTheory}
 &&p_n(t) = \dot u_n \approx A(n,t)B(n,t), \\
  &&A(n,t)=\omega\sqrt{h_3} \cn \Big[\sqrt{h_3-h_1}\sqrt{\frac{3\beta \omega^2}{2k^2}}\Big(n -v t\Big)-C_3,m\Big], \nonumber\\
 &&B(n,t) = \sin\Big[qn-\Big(\omega+\frac{3\beta \omega^3}{16k^2}\big(h_3+h_1\big)\Big)t -C_4 \Big]. \nonumber
\end{eqnarray}

\begin{figure}
\includegraphics[width=\linewidth]{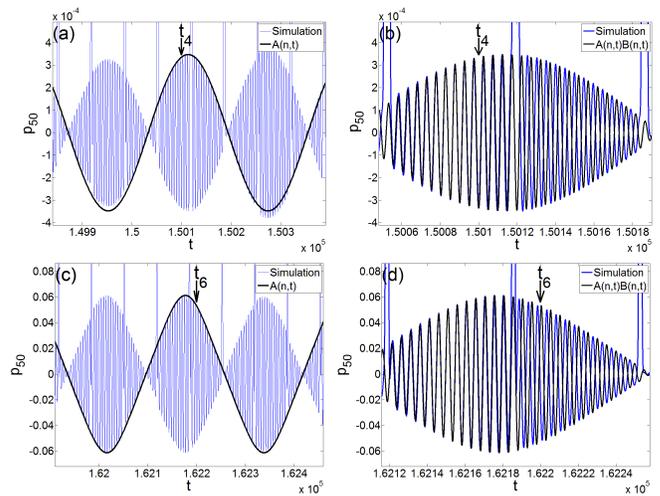}
\caption{\label{MomentumFit} (Color online) Blue lines: time evolution data $p_{50}(t)$ in the neighborhoods of $t_4$ and $t_6$. Black lines in left panels: $A(n,t)$ of the analytic formula (\ref{ParticleMomentumTheory}). Black lines in right panels: the analytic formula (\ref{ParticleMomentumTheory}).}
\end{figure}

Although the solitary wave is highly localized in space, it has an infinite span.
When the solitary wave is restricted to a ring with finite size, infinitesimal perturbations will appear.
Due to the modulational instability of the NLS equation~(\ref{NLSEquationFromFPUbeta})~\cite{Kivshar2000}, the infinitesimal perturbation grows as shown in Figs.~\ref{Momentum}(a) and~\ref{Momentum}(b).
Since a more energetic packet samples the more anharmonic portions of the potential, one expects nonlinearity will play a much more important role as the magnitude of the low-amplitude wave increases.
After a critical point, the low-amplitude wave may spontaneously self-modulate and split into "wave packets" as shown in Fig.~\ref{Momentum}(c).
These wave packets may coexist with the solitary wave associated with the periodic exchange of energy between them as indicated by the periodic oscillation of $E_W$ in the left-hand side of Fig.~\ref{LAWEnergy}(b). Figures~\ref{Momentum}(c)-\ref{Momentum}(f) present the transfer of the energy from the solitary wave to these wave packets.

We shall here verify that the low-amplitude wave packets after the self-modulation is the cnoidal wave described by our analytic formula~(\ref{ParticleMomentumTheory}).
The time evolution data of a single particle is recorded ($p_{50}$ in the present work) and its low-amplitude part is compared with Eq.~(\ref{ParticleMomentumTheory}) as follows.
Note that $A(n,t)$ is a periodic function of $n$ and $t$ with periods $T_n = \frac{4K(m)}{\sqrt{h_3-h_1}} \sqrt{\frac{2k^2}{3\beta \omega^2}}$ and $T_t = \frac{1}{v} \frac{4K(m)}{\sqrt{h_3-h_1}} \sqrt{\frac{2k^2}{3\beta \omega^2}}$ respectively, where $K(m)=\int_0^{\pi/2} \frac{d \vartheta}{\sqrt{1-m \sin^2\vartheta}}$ is the complete elliptic integral of the first kind.
According to the time evolution data $p_{50}(t)$, both $h_3$ and $m$ as functions of time can be obtained. We present $m$ in the right-hand side of Fig.~\ref{LAWEnergy}(b). Note that both $h_3$ and $m$ are slow variables and keep invariant approximatively in a relatively short time scale.
The low-amplitude part of $p_{50}(t)$ and the cnoidal wave described by Eq.~(\ref{ParticleMomentumTheory}) are plotted in Fig.~\ref{MomentumFit} for two typical time ranges, i.e., the neighborhood of $t_4$ as shown in Figs.~\ref{MomentumFit}(a) and~\ref{MomentumFit}(b) (linear limit with $m \approx 2.3094\times 10^{-5}$) and the neighborhood of $t_6$ as shown in Figs.~\ref{MomentumFit}(c) and~\ref{MomentumFit}(d) (strong nonlinearity with $m \approx 0.51181$).
It is clearly seen that the low-amplitude wave agrees well with the cnoidal wave. The interaction of the cnoidal wave with the solitary wave has two consequences, i.e., the slow variation of the amplitude as shown clearly in Fig.~\ref{MomentumFit}(a) and a spatial shift as shown in Figs.~\ref{MomentumFit}(b) and~\ref{MomentumFit}(d).

\begin{figure}
\includegraphics[width=\linewidth]{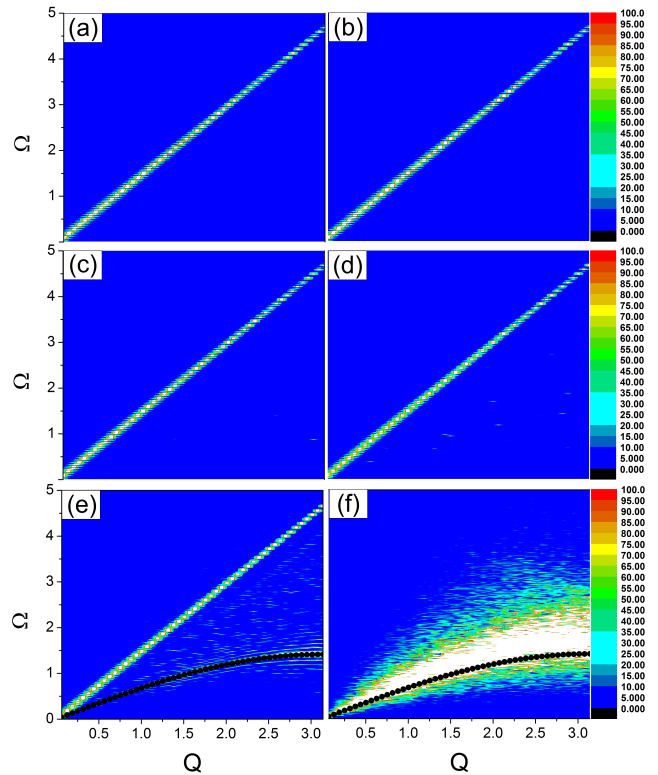}
\caption{\label{Dispersion} (Color online) Contour plot of the SED for the FPU-$\beta$ ring. (a)-(f) correspond to $R_1$-$R_6$ indicated in Fig.~\ref{LAWEnergy}(a), respectively. The dark dots in (e) and (f) stand for the phonon dispersion relation of the corresponding harmonic lattice.}
\end{figure}

\begin{figure}
\includegraphics[width=\linewidth]{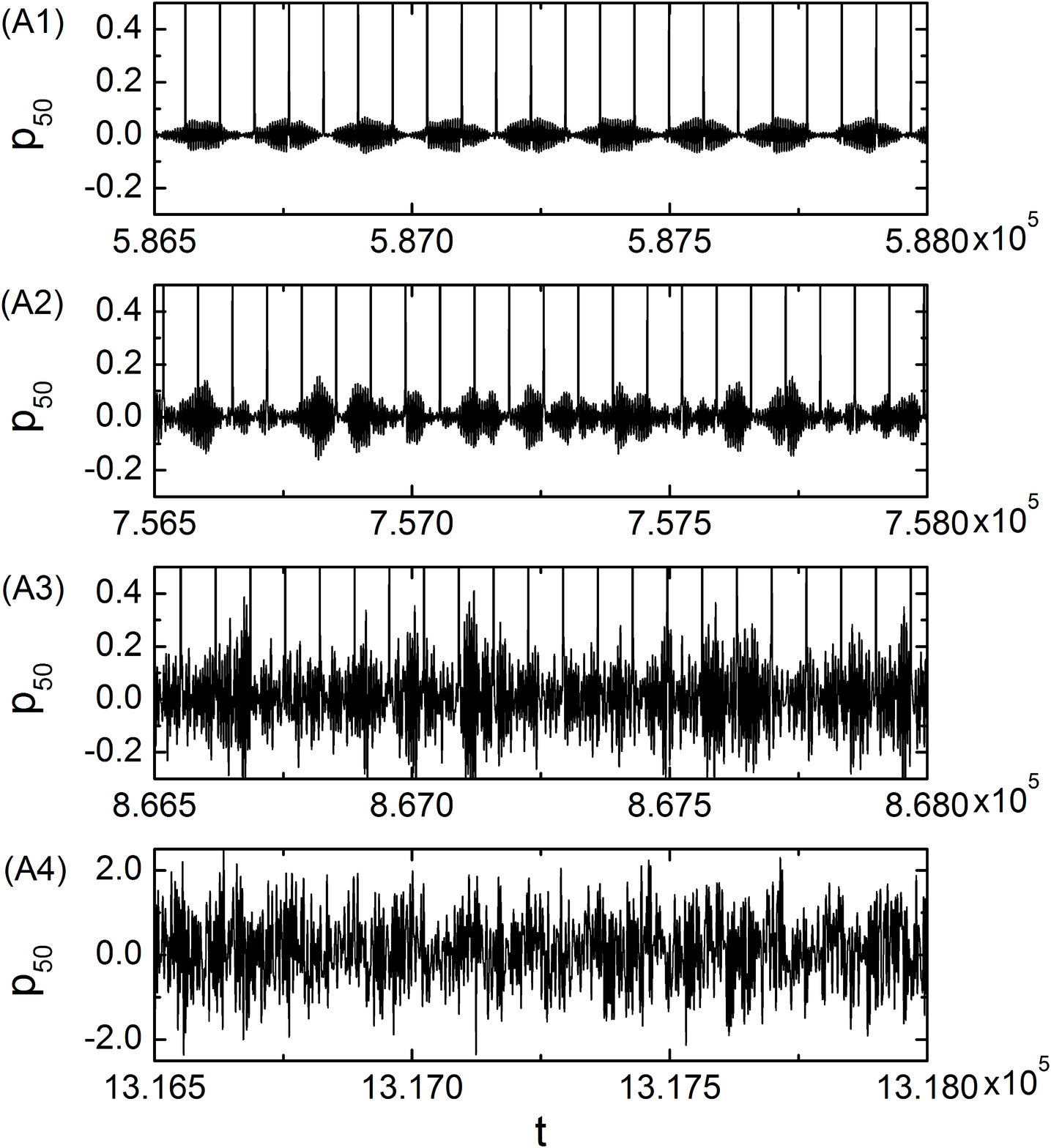}
\caption{\label{TimeEvolutionPart} (A1)-(A4): parts of the time evolution data $p_{50}(t)$ of the time regions $R_3$-$R_6$ indicated in Fig.~\ref{LAWEnergy}(a).}
\end{figure}

Till now, we have focused on the first stage of the process of the loss of stability of a solitary wave on a FPU ring where a cnoidal wave is excited and coexists with the solitary wave.
To get a deep understanding of the following stages of the process, the SED method is employed to predict the dispersion relations. Figure~\ref{Dispersion} presents the SED for the FPU-$\beta$ ring in different time regions $R_1$-$R_6$ indicated in Fig.~\ref{LAWEnergy}(a).
The shading on the plot stands for the magnitude of the SED for each $(Q,\Omega)$ combination corresponding to a total integration time $2.5\times 10^4$ with sample interval $0.01$.
It is clear that there are no phonons in the time regions $R_1$ and $R_2$ which correspond to the first stage of the process. The straight lines in Figs.~\ref{Dispersion}(a) and~\ref{Dispersion}(b) stand for the solitary wave and their slop is consistent with the velocity of the solitary wave.
In the second stage of the process, the interaction of the cnoidal wave with the solitary wave starts to take effect and radiate phonons (see Figs.~\ref{Dispersion}(c) and~\ref{Dispersion}(d)), by which the cnoidal wave is deformed and destroyed as shown clearly by parts of the time evolution data $p_{50}(t)$ of the time regions $R_3$ and $R_4$ (see Figs.~\ref{TimeEvolutionPart}(A1) and~\ref{TimeEvolutionPart}(A2)).
In the third stage of the process, the interaction between phonons and the solitary wave leads to the radiation of more phonons and the collapse of the solitary wave as shown in Figs.~\ref{Dispersion}(e) and~\ref{Dispersion}(f).
Figures~\ref{TimeEvolutionPart}(A3) and~\ref{TimeEvolutionPart}(A4) present parts of the time evolution data $p_{50}(t)$ of the time regions $R_5$ and $R_6$. No trace of solitary wave could be found in Fig.~\ref{TimeEvolutionPart}(A4).
In Figs.~\ref{Dispersion}(e) and~\ref{Dispersion}(f), the dark dots stand for the phonon dispersion relation of the corresponding harmonic lattice.
Due to the nonlinearity, the SED for the FPU-$\beta$ ring in the equilibrium state shifts toward higher frequencies compared with the harmonic case as shown in Fig.~\ref{Dispersion}(f).

\begin{figure}
\includegraphics[width=\linewidth]{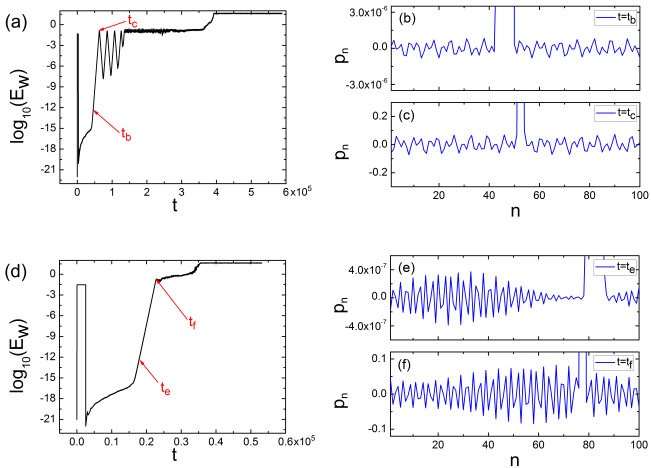}
\caption{\label{Discussion} (Color online) (a) and (d) Time evolution of $E_W$ for $\beta=0.335$ and $\beta=0.435$. (b), (c) and (e), (f) Momentum distribution profiles among the particles for $t_b$, $t_c$ and $t_e$, $t_f$ as indicated in (a) and (d).}
\end{figure}

Nonlinear Hamiltonian systems may exhibit distinctly different dynamical behaviors for different energy levels.
Since increasing the nonlinear parameter $\beta$ is equivalent to increasing the energy in our model~\cite{Aoki2001}, we investigated the propagation behavior of the solitary wave for $\beta \in [0.1,1]$ with sample interval $\Delta \beta =0.001$.
Figure \ref{Discussion}(a) presents the time evolution of $E_W$ for $\beta=0.335$ and the momentum distribution profiles among the particles for $t=t_b$ and $t=t_c$ are depicted in Figs. \ref{Discussion}(b) and \ref{Discussion}(c), respectively. Here, owing to the modulational instability of Eq.~(\ref{NLSEquationFromFPUbeta}), the formation of the low-amplitude wave can be clearly observed. Although the cyclic energy exchange between the solitary and low-amplitude waves is found to be generic for ¦Â$\beta \in [0.1,0.4]$, the low-amplitude waves may be complicated, rather than a simple cnoidal wave, where further studying is needed.
When $\beta$ is further increased, the cyclic energy exchange is rare and the solitary wave loses stability rapidly after the formation of the low-amplitude (cnoidal or complicated) wave. One typical example is shown in Figs. \ref{Discussion}(d)-\ref{Discussion}(f) for $\beta=0.435$. We explained that the system has entered the strong chaos regime and become very unstable~\cite{Casetti1995,Casetti1996,Yuan2013}.

\section{conclusion and discussions}
In summary, one possible mechanism of the loss of stability of a solitary wave on a FPU-$\beta$ ring is revealed.
We observed numerically the coexistence of a solitary wave with a cnoidal wave associated with the periodic exchange of energy between these two nonlinear waves.
Due to the interaction of the cnoidal wave with the solitary wave, phonons can be radiated, which destroy the cnoidal wave and finally result in the loss of stability of the solitary wave.
For some values of $\beta$ in our FPU ring, the coexistence of the solitary wave and the cnoidal wave has no tendency to disappear up to $1 \times 10^8$ time unit (although the energy exchange may become quasi-periodic as time evolves).
We should emphasize that these cases may correspond to the situations where linear unstable solitary waves are actually stable in the nonlinear sense~\cite{Berryman1976}, and a more careful investigation is still needed.

In the framework of the one-dimensional NLS equation, the interaction of solitons (envelope solitons) with radiation (a nonsoliton part)~\cite{Kuznetsov1995}, continuous waves having constant amplitude~\cite{Park1999}, cnoidal waves~\cite{Shin2001}, and continuous waves of arbitrary shape~\cite{Shin2003} has been studied both analytically and numerically. The bright solitons on a cnodial wave background train of the inhomogeneous NLS equation have also been investigated~\cite{Murali2008}.
The theoretical analysis of the interaction between the solitons and the cnoidal waves in the NLS equation (integrable) is based on the construction of the exactly superposed solutions of these two waves. It is reported that the solitons can recover their original shapes and velocities after collisions, while shapes of cnoidal waves are nearly preserved during collisions~\cite{Shin2001}.
In the FPU lattices, which are nonintegrable, however, the interaction between the solitary waves and the cnoidal waves is untouched before, to our knowledge. In the present paper, according to our simulation results, the interaction between these two nonlinear waves results in the energy exchange and the radiation of phonons.
These findings might be helpful to find an appropriate approximation method to investigate the interaction theoretically.

In regard to the experimental observation and potential practical applications, we notice the nonlinear oscillations of a liquid drop.
A liquid drop possessed a circular geometry in nature, and the studies of its free oscillations have a long history~\cite{Trinh1982,Tian1995,Apfel1997,Ludu1997}.
The traveling waves on liquid drops are observed~\cite{Trinh1988} and related to traveling deformations called "rotons"~\cite{Ludu1998}. These rotons can range from small oscillations (linearized model), to cnoidal waves, and on out to solitary waves~\cite{Ludu1998}.
We hope that our results in the FPU ring might find applications in the fundamental study of the phenomenon in the liquid drops and also droplike systems, i.e., astronomical objects~\cite{Smarr1973,Cardoso2006,Lacerda2007} and atomic nuclei~\cite{Gherghescu2001,Schunck2007}.

\begin{acknowledgments}
We thank Jie Liu and Li-Bin Fu at the Institute of Applied Physics and Computational Mathematics for enlightening discussions. Z.Y., M.C. and Z.Z. were supported partly by the National Natural Science Foundation of China (Grant No. 11075016) and the Research Fund for the Doctoral Program of Higher Education of China (Grant No. 20100003110007). J.W. and G.X. were supported partly by the National Basic Research Program of China (Program No. 2011CB710704).
\end{acknowledgments}

\bibliography{References}

\end{document}